\documentclass[preprint2]{aastex63}
\usepackage{float, bm, graphicx, amsmath, morefloats}
\usepackage{booktabs}
\usepackage{CJK,CJKspace,CJKpunct}
\usepackage{color}

\submitjournal{ApJ}

\shorttitle{SN\,2018bvw}
\shortauthors{Ho et al.}

\newcommand{\name}{ZTF18aaqjovh}

\newcommand{\msol}{\mbox{$\rm M_\odot$}}
\newcommand{\mni}{\mbox{$\rm M_{Ni}$}}
\newcommand{\mej}{\mbox{$\rm M_{ej}$}}

\newcommand{\days}{\mbox{$\rm days$}}
\newcommand{\psec}{\mbox{$\rm s^{-1}$}}
\newcommand{\pyr}{\mbox{$\rm yr^{-1}$}}

\newcommand{\degsq}{\mbox{$\rm deg^{2}$}}

\newcommand{\km}{\mbox{$\rm km$}}
\newcommand{\pcmsq}{\mbox{$\rm cm^{-2}$}}
\newcommand{\mpc}{\mbox{$\rm Mpc$}}

\newcommand{\kev}{\mbox{$\rm keV$}}
\newcommand{\erg}{\mbox{$\rm erg$}}

\newcommand{\ghz}{\mbox{$\rm GHz$}}
\newcommand{\phz}{\mbox{$\rm Hz^{-1}$}}

\newcommand{\counts}{\mbox{$\rm ct$}}

\newcommand{\nickel}{\mbox{$\rm {}^{56}Ni$}}
\newcommand{\cobalt}{\mbox{$\rm {}^{56}Co$}}

\begin{document}

\title{The Broad-lined Ic Supernova ZTF18aaqjovh (SN\,2018bvw): \\ An Optically Discovered Engine-driven Supernova Candidate with Luminous Radio Emission}

\author[0000-0002-9017-3567]{Anna Y. Q.~Ho}
\affiliation{Cahill Center for Astrophysics, 
California Institute of Technology, MC 249-17, 
1200 E California Boulevard, Pasadena, CA, 91125, USA}

\author{Alessandra Corsi}
\affiliation{Department of Physics and Astronomy,
Texas Tech University,
Box 1051, Lubbock, TX 79409-1051, USA}

\author{S. Bradley~Cenko}
\affiliation{Astrophysics Science Division, 
NASA Goddard Space Flight Center, 
Mail Code 661, Greenbelt, MD 20771, USA}
\affiliation{Joint Space-Science Institute, University of Maryland, College Park, MD 20742, USA}

\author{Francesco Taddia}
\affiliation{Department of Astronomy, The Oskar Klein Center, Stockholm University, AlbaNova, 10691 Stockholm, Sweden}

\author{S. R.~Kulkarni}
\affiliation{Cahill Center for Astrophysics, 
California Institute of Technology, MC 249-17, 
1200 E California Boulevard, Pasadena, CA, 91125, USA}

\author{Scott Adams}
\affiliation{Cahill Center for Astrophysics, 
California Institute of Technology, MC 249-17, 
1200 E California Boulevard, Pasadena, CA, 91125, USA}

\author{Kishalay De}
\affiliation{Cahill Center for Astrophysics, 
California Institute of Technology, MC 249-17, 
1200 E California Boulevard, Pasadena, CA, 91125, USA}

\author{Richard Dekany}
\affiliation{Caltech Optical Observatories, California Institute of Technology, Pasadena, CA, USA}

\author[0000-0002-1153-6340]{Dmitry D. Frederiks}
\affiliation{Ioffe Institute, Politekhnicheskaya 26, St. Petersburg 194021, Russia}

\author{Christoffer Fremling}
\affiliation{Cahill Center for Astrophysics, 
California Institute of Technology, MC 249-17, 
1200 E California Boulevard, Pasadena, CA, 91125, USA}

\author[0000-0001-8205-2506]{V. Zach Golkhou}
\affiliation{DIRAC Institute, Department of Astronomy, University of Washington, 3910 15th Avenue NE, Seattle, WA 98195, USA} 
\affiliation{The eScience Institute, University of Washington, Seattle, WA 98195, USA}
\altaffiliation{Moore-Sloan, WRF Innovation in Data Science, and DIRAC Fellow}

\author[0000-0002-3168-0139]{Matthew J.\ Graham}
\affiliation{Cahill Center for Astrophysics, 
California Institute of Technology, MC 249-17, 
1200 E California Boulevard, Pasadena, CA, 91125, USA}

\author[0000-0002-9878-7889]{Tiara Hung}
\affiliation{Department of Astronomy and Astrophysics,
University of California, Santa Cruz, California, 95064, USA}

\author[0000-0002-6540-1484]{Thomas Kupfer}
\affiliation{Kavli Institute for Theoretical Physics, University of California, Santa Barbara, CA 93106, USA}

\author[0000-0003-2451-5482]{Russ R. Laher}
\affiliation{IPAC, California Institute of Technology, 1200 E. California Boulevard, Pasadena, CA 91125, USA}
             
\author{Ashish Mahabal}
\affiliation{Cahill Center for Astrophysics, 
California Institute of Technology, MC 249-17, 
1200 E California Boulevard, Pasadena, CA, 91125, USA}
\affiliation{Center for Data Driven Discovery, California Institute of Technology, Pasadena, CA 91125, USA}

\author[0000-0002-8532-9395]{Frank J. Masci}
\affiliation{IPAC, California Institute of Technology, 1200 E. California Blvd, Pasadena, CA 91125, USA}

\author[0000-0001-9515-478X]{Adam A. Miller}
\affiliation{Center for Interdisciplinary Exploration and Research in Astrophysics (CIERA) and Department of Physics and Astronomy, Northwestern University, 2145 Sheridan Road, Evanston, IL 60208, USA}
\affiliation{The Adler Planetarium, Chicago, IL 60605, USA}

\author[0000-0002-0466-1119]{James D. Neill}
\affiliation{Caltech Optical Observatories, California Institute of Technology, Pasadena, CA, USA}

\author{Daniel Reiley}
\affiliation{Caltech Optical Observatories, California Institute of Technology, Pasadena, CA, USA}

\author[0000-0002-0387-370X]{Reed Riddle}
\affiliation{Caltech Optical Observatories, California Institute of Technology, Pasadena, CA, USA}

\author{Anna Ridnaia}
\affiliation{Ioffe Institute, Politekhnicheskaya 26, St. Petersburg 194021, Russia}

\author[0000-0001-7648-4142]{Ben Rusholme}
\affiliation{IPAC, California Institute of Technology, 1200 E. California Blvd, Pasadena, CA 91125, USA}

\author{Yashvi Sharma}
\affiliation{Cahill Center for Astrophysics, 
California Institute of Technology, MC 249-17, 
1200 E California Boulevard, Pasadena, CA, 91125, USA}

\author{Jesper Sollerman}
\affiliation{Department of Astronomy, The Oskar Klein Center, Stockholm University, AlbaNova, 10691 Stockholm, Sweden}

\author[0000-0001-6753-1488]{Maayane T. Soumagnac}
\affiliation{Lawrence Berkeley National Laboratory, 1 Cyclotron Road, Berkeley, CA 94720, USA}
\affiliation{Department of Particle Physics and Astrophysics, Weizmann Institute of Science, Rehovot 76100, Israel}

\author[0000-0002-2208-2196]{Dmitry S. Svinkin}
\affiliation{Ioffe Institute, Politekhnicheskaya 26, St. Petersburg 194021, Russia}

\author[0000-0003-4401-0430]{David L. Shupe}
\affiliation{IPAC, California Institute of Technology, 1200 E. California Blvd, Pasadena, CA 91125, USA}

\correspondingauthor{Anna Y. Q. Ho}
\email{ah@astro.caltech.edu}

\received{19 Dec 2019}
\revised{26 Feb 2020}
\accepted{2 Mar 2020}

\submitjournal{The Astrophysical Journal}

\begin{abstract}

We present \name\ (SN\,2018bvw), a
high-velocity (``broad-lined'') stripped-envelope (Type~Ic) supernova (Ic-BL SN)
discovered in the Zwicky Transient Facility
one-day cadence survey.
\name\ shares a number of features in common with engine-driven explosions:
the photospheric velocity and the shape of the optical light curve
are very similar to that of the Type~Ic-BL SN\,1998bw,
which was associated with a low-luminosity gamma-ray burst (LLGRB) and had relativistic ejecta.
However, the radio luminosity of \name\ is almost two orders of magnitude fainter than that of SN\,1998bw at the same velocity phase,
and the shock velocity is at most mildly relativistic ($v = 0.06$--0.4$c$).
A search of high-energy catalogs reveals no compelling gamma-ray burst (GRB) counterpart to \name,
and the limit on the prompt GRB luminosity
of $L_{\gamma,\mathrm{iso}} \approx 1.6 \times 10^{48}\,\erg\,\psec$ excludes a classical GRB but not an LLGRB.
Altogether, \name\ represents another transition event between engine-driven SNe associated with GRBs and ``ordinary'' Ic-BL SNe.

\end{abstract}

\section{Introduction}
\label{sec:introduction}

Broad-lined Type Ic supernovae (Ic-BL SNe)
are a subclass of stripped-envelope core-collapse supernovae (CC SNe) characterized by fast ejecta and large kinetic energies.
While typical Type Ic SNe have photospheric velocities
$v_\mathrm{ph}\approx10,000\,\km\,\psec$ (measured from \ion{Fe}{2} absorption features),
Type Ic-BL SNe have $v_\mathrm{ph}\approx20,000\,\km\,\psec$ at maximum light \citep{Modjaz2016}.
The kinetic energy release of Ic-BL SNe is typically $\sim10^{52}\,\erg$ \citep{Cano2013,Lyman2016,Prentice2016}, an order of magnitude greater than traditional CC SNe \citep{Woosley2005}, although this measurement is highly model-dependent.

A clue to the high energies and fast velocities present in Ic-BL SNe is their connection to long-duration gamma-ray bursts (GRB),
reviewed in \citet{Woosley2006} , \citet{Hjorth2012}, and \citet{Cano2017}.
The association began with the
coincident discovery of GRB\,980425 and SN\,1998bw at $d = 40\,$Mpc \citep{Galama1998,Kulkarni1998}.
However, GRB\,980425 was different from typical GRBs:
it was underluminous in $\gamma$-rays
($L_{\gamma,\mathrm{iso}} \sim 5 \times 10^{46}\,\erg\,\psec$
compared to typical values of $10^{51}$--$10^{53}\,\erg\,\psec$)
and subenergetic, with an isotropic equivalent energy four orders of magnitude smaller than that of typical GRBs.
Thus, it took the discovery of the cosmological GRB\,030329 ($z = 0.1685$)
in association with SN\,2003dh \citep{Hjorth2003,Stanek2003}
to solidify the relationship between GRBs and SNe.

Since then, $\sim20$ SNe accompanying GRBs have been spectroscopically confirmed. All show broad Type~Ic-BL features near maximum light, with two exceptions: SN\,2011kl had a relatively featureless spectrum, and SN\,2013ez more closely resembled a Type~Ic \citep{Cano2017}.
The GRB-SN association has led to the
suggestion that GRBs and Ic-BL SNe are powered by a single central engine \citep{Lazzati2012,Sobacchi2017,Barnes2018}.
However, a systematic search for radio emission from Ic-BL SNe constrained the fraction harboring a relativistic outflow as bright as that of SN\,1998bw to be $\lesssim14\%$ \citep{Corsi2016}.

Complicating matters, additional underluminous GRBs have been discovered since GRB\,980425
and are collectively referred to as
low-luminosity GRBs (LLGRBs). LLGRBs are distinguished by isotropic 
peak luminosities $L_\mathrm{iso} \approx 10^{46}$--$10^{48} \,\erg\,\psec$
and a relativistic energy release that is 2--3 orders of magnitude
smaller than the $10^{51}\,\erg$ from GRBs with fully relativistic outflows \citep{Cano2017}.
Due to their lower intrinsic luminosities,
LLGRBs are discovered at low redshifts ($z \lesssim 0.1$).
Thus, despite the fact that their intrinsic rate might be 10--100 higher than that of classical GRBs \citep{Soderberg2006,Pian2006},
only seven have been discovered:
LLGRB\,980425/SN\,1998bw, 
XRF\,020903 \citep{Sakamoto2004,Soderberg2004a,Bersier2006},
LLGRB\,031203/SN\,2003lw \citep{Malesani2004,Soderberg2004b,Thomsen2004,Watson2004},
LLGRB\,060218/SN\,2006aj \citep{Mirabal2006,Pian2006,Soderberg2006}, 
LLGRB\,100316D/SN\,2010bh \citep{Starling2011,Bufano2012},
LLGRB\,171205A/SN\,2017iuk \citep{DElia2018,Wang2018},
and most recently LLGRB\,190829A \citep{Chand2020}.
LLGRB\,060218 and LLGRB\,100316D
have their own distinct properties:
a long $\gamma$-ray prompt emission phase,
and long-lived soft X-ray emission that might
arise from continued activity of the central engine \citep{Soderberg2006,Margutti2013} or dust echoes \citep{Margutti2015,Irwin2016}.

Modeling of the radio emission from LLGRBs suggests quasi-spherical ejecta coupled to mildly relativistic material, with no off-axis components \citep{Kulkarni1998,Soderberg2006,Pian2006,Margutti2013}.
Thus, it seems that LLGRBs arise from a fundamentally different mechanism to cosmological GRBs.
One suggestion is that they represent failed or choked-jet events, and that the gamma rays arise from shock breakout.
This is supported by the early light curve of the LLGRB 060218,
whose double peak in ultraviolet and optical filters has been
modeled as shock breakout into a dense stellar wind \citep{Campana2006} or into an extended envelope \citep{Nakar2015}.
Another possibility is that the prompt emission is from a successful low-luminosity jet \citep{Irwin2016}.

A major focus of scientific investigation over the past 20 years has been to unify this diverse array of phenomena: ``extreme'' SNe with successful, observed jets (classical GRBs), mildly relativistic explosions (LLGRBs or radio-emitting SNe), and ordinary (nonrelativistic) SNe.
The traditional avenue to discovering central engines
-- the detection of a GRB -- is severely limited
because a number of conditions must be met for a central engine to produce a GRB. First, the jet must be nearly baryon-free---else the available energy is insufficient to accelerate the ejecta to ultrarelativistic velocities, and gamma-ray emission will be stifled by pair-production \citep{Piran2004}. Next, the jet must successfully escape the star without being choked by the stellar envelope \citep{MacFadyen2001}. Finally, the jet must be directed at Earth.

Today, wide-field optical time-domain surveys have the field of view and cadence to discover engine-driven explosions without relying on a high-energy trigger (e.g. \citealt{Corsi2017}).
Radio observations are central to this effort,
because they trace the fastest-moving ejecta.
The Zwicky Transient Facility (ZTF; \citealt{Bellm2019a,Graham2019}) is conducting several different surveys \citep{Bellm2019b}
using a custom mosaic camera \citep{Dekany2016} on the 48 inch Samuel Oschin Telescope (P48) at the Palomar Observatory.
ZTF discovers one Ic-BL SN per month,
and we are conducting a follow-up campaign of a subset of these events with the Karl G. Jansky Very Large Array (VLA; \citealt{Perley2011}). Here, we present our first detection of radio emission from the Ic-BL \name\ (SN\,2018bvw).
In Section \ref{sec:observations} we describe our optical, radio, and X-ray observations, as well as our search for contemporaneous gamma-ray emission.
In Section 3 we constrain the physical properties of the explosion (energy, velocity, and ejecta mass).
We present our conclusions in Section 4.

Throughout the paper we use the $\Lambda$CDM cosmology from \citet{Planck2016}.

\section{Observations}
\label{sec:observations}

\subsection{Zwicky Transient Facility Discovery}
\label{sec:discovery}

ZTF images are processed and reference-subtracted by the IPAC ZTF pipeline \citep{Masci2019} using the method described in \citet{Zackay2016},
and every 5$\sigma$ point-source detection is saved as an ``alert.''
Alerts are distributed in Apache Avro format \citep{Patterson2019} and can be filtered based on a machine-learning real-bogus metric \citep{Duev2019,Mahabal2019}, host characteristics (including a star-galaxy classifier; \citealt{Tachibana2018}\footnote{In this context TM18 define star as an unresolved point source and galaxy as an extended unresolved source}), and light-curve properties.
The ZTF collaboration uses a web-based system called the GROWTH marshal \citep{Kasliwal2019} to identify, monitor, and coordinate follow-up observations for transients of interest.

\name\ was discovered in an image obtained on 2018 May 5 UT as part of the ZTF one-day cadence survey, which covers 3000\,\degsq\ in two visits (one $g$, one $r$) per night \citep{Bellm2019b}.
The alert passed two filters, as part of two systematic surveys being conducted by ZTF:
a filter for transients in the local universe that cross-matches sources with a catalog of nearby galaxies
\citep{Cook2019},
and a filter for bright transients \citep{Fremling2019}.
Because it passed these filters,
the source was reported to the Transient Name Server (TNS\footnote{https://wis-tns.weizmann.ac.il}; \citealt{Fremling2018}) and received the designation SN\,2018bvw.
After being reported, it was
spectroscopically classified (Section \ref{sec:spectra}; \citealt{Fremling2019classification}).

\begin{figure}[!htb]
\centering
\includegraphics[scale=0.6]{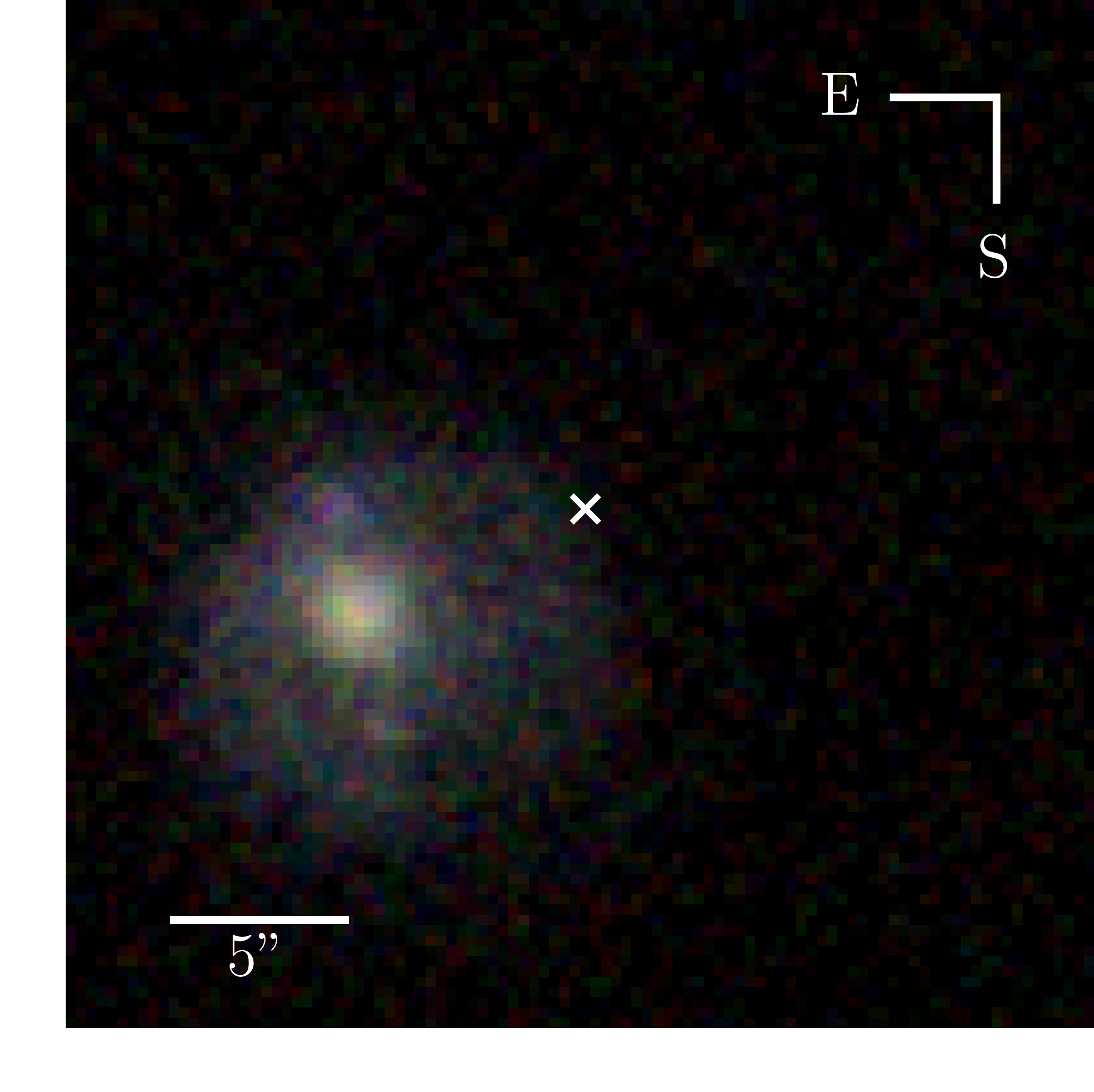}
\caption{Image of the host galaxy of \name\ (SN\,2018bvw), constructed from $g$-, $r$-, and $i$-band SDSS cutouts. The position of \name\ is shown with a white cross,
$4.71\arcsec$ from the center of the galaxy, or $5.68$\,kpc assuming $d=249\,\mpc$.}
\label{fig:cutout}
\end{figure}

The discovery magnitude was $r=18.65 \pm 0.02$ mag,
where the error bar is a 1$\sigma$ estimate of the background RMS, derived using a pixel-uncertainty map created for the difference image \citep{Masci2019}.
The source position was measured to be
$\alpha = 11^{\mathrm{h}}52^{\mathrm{m}}43.62^{\mathrm{s}}$, $\delta = +25^{\mathrm{d}}40^{\mathrm{m}}30.1^{\mathrm{s}}$ (J2000).
The position is 4.71\arcsec\ from SDSS J115244.11+254027.1, a star-forming galaxy at $z=0.05403 \pm 0.00001$ (248.85\,\mpc; \citealt{Alam2015}).
The transient position with respect to the host galaxy is shown in Figure \ref{fig:cutout},
with the host galaxy image constructed from
SDSS $g$-, $r$-, and $i$-band cutouts using the method in \citet{Lupton2004}. 
At this distance, the projected offset between \name\ and the center of the host corresponds to $d=5.68$\,kpc.
This offset is larger than the typical offset of Ic-BL SNe accompanied by GRBs, which is $1.54^{+3.13}_{-1.28}$\,kpc (1$\sigma$ confidence),
and more consistent with the offsets of Ic-BL SNe without detected GRBs,
measured to be $\left(3.08^{+2.98}_{-2.35}\right)$\,kpc \citep{Japelj2018}.

\begin{figure}[!htb]
\centering
\includegraphics[width=\columnwidth]{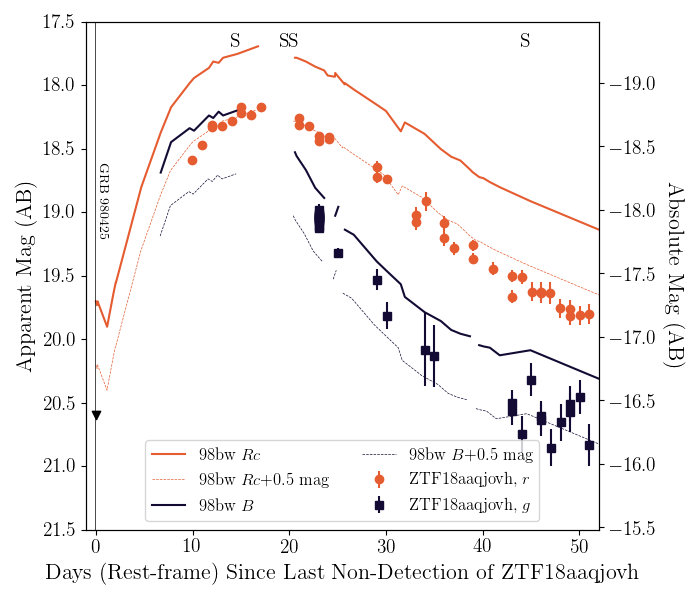}
\caption{Optical light curve of \name, corrected for Milky Way extinction,
with P48 $r$-band in orange circles and P48 $g$-band in black squares.
The light curve of SN\,1998bw from Table 2 of \citet{Clocchiatti2011} is shown for comparison as thick black ($B$-band) and thick orange ($Rc$-band) lines, shifted to the redshift of \name\ and also corrected for Milky Way extinction.
The same SN\,1998bw light curves are shifted by 0.4\,mag for closer comparison and are shown as thin dotted lines.
The vertical line on the left-hand side indicates the relative time of the GRB\,980425, the low-luminosity gamma-ray burst associated with SN\,1998bw.
The epochs of optical spectra of \name\ are marked with `S' along the top of the figure.}
\label{fig:lc}
\end{figure}

The full light curve, corrected for Milky Way extinction, is provided in Table \ref{tab:opt-phot} and shown in Figure \ref{fig:lc}.
The P48 measurements come from forced photometry \citep{Yao2019}.
The $g$-band reference image was constructed from data taken between 2018 April 22 and 2018 May 16, so we had to subtract a baseline flux to account for SN light in the reference.
To calculate the baseline flux, we measured the mean flux of photometry in images where the SN light was not present: a set of images at $\Delta t\approx\,-50\days$ and a set of images at $\Delta t \approx 400\,\days$.
We confirmed that this baseline level was consistent,
i.e. that by 400\,\days\ the SN light had returned to a level consistent with the pre-explosion level.

We obtained two epochs of photometry from the Spectral Energy Distribution Machine (SEDM; \citealt{Blagorodnova2018,Rigault2019}) mounted on the automated 60 inch telescope at Palomar (P60; \citealt{Cenko2006}).
Digital image subtraction and photometry for the SEDM was performed using the Fremling Automated Pipeline (\texttt{FPipe}; \citealt{Fremling2016}).
\texttt{Fpipe} performs calibration and host subtraction against SDSS reference images and catalogs \citep{Ahn2014}.

 The peak $r$-band absolute magnitude is typical of Ic-BL light curves compiled from untargeted surveys \citep{Taddia2019},
 and the light curve of \name\ is very similar in shape to the light curve of SN\,1998bw (Figure \ref{fig:lc}).
 Assuming that the time from explosion to peak is the same in \name\ as in SN\,1998bw, we can estimate that the explosion time $t_0$ is about the time of the last nondetection, 2018 April 25 UT.
 The optical spectra of \name\
 (Section \ref{sec:spectra}) suggest that this $t_0$ is accurate to within a few days: the spectrum of \name\ on May 9 was most similar to that of SN\,1998bw at 16\,\days\ post-explosion. With this $t_0$,
 the first detection of \name\ by ZTF was at $\Delta t=10$ days.
 Throughout the paper, we use this definition of
 $t_0$ and report all times $\Delta t$ with respect to this reference point.

\startlongtable
\begin{deluxetable}{lrrrrr}
\tablecaption{Optical light curve of ZTF18aaqjovh. P48 values are from forced photometry \citep{Yao2019}.\label{tab:opt-phot}} 
\tablewidth{0pt} 
\tablehead{ \colhead{Date} & \colhead{$\Delta t$} & \colhead{Instr.} & \colhead{Filt.} & \colhead{Mag}\\
\colhead{(MJD)} & \colhead{(days)} & & & \colhead{(AB)}
} 
\tabletypesize{\scriptsize} 
\startdata 
58217.239572 & -15.94 & P48 & $r$ & $22.99 \pm 1.13$ \\ 
58217.240521 & -15.94 & P48 & $r$ & $24.54 \pm 4.88$ \\ 
58217.262072 & -15.91 & P48 & $r$ & $23.49 \pm 3.69$ \\ 
58218.257859 & -14.92 & P48 & $r$ & $22.26 \pm 0.75$ \\ 
58219.216238 & -13.96 & P48 & $r$ & $22.99 \pm 1.52$ \\ 
58219.262523 & -13.91 & P48 & $r$ & $23.57 \pm 2.56$ \\ 
58221.255984 & -11.92 & P48 & $r$ & $21.59 \pm 0.80$ \\ 
58222.259039 & -10.92 & P48 & $r$ & $22.95 \pm 1.78$ \\ 
58224.218553 & -8.96 & P48 & $r$ & $22.19 \pm 1.30$ \\ 
58224.226030 & -8.95 & P48 & $r$ & $22.83 \pm 1.62$ \\ 
58224.233669 & -8.94 & P48 & $r$ & $23.99 \pm 4.17$ \\ 
58224.324641 & -8.85 & P48 & $r$ & $20.87 \pm 0.67$ \\ 
58227.250903 & -5.93 & P48 & $r$ & $23.10 \pm 1.64$ \\ 
58231.276227 & -1.90 & P48 & $r$ & $23.96 \pm 4.22$ \\ 
58233.175208 & -0.00 & P48 & $r$ & $24.15 \pm 6.19$ \\ 
58233.176146 & 0.00 & P48 & $r$ & $22.73 \pm 1.79$ \\ 
58243.170324 & 9.99 & P48 & $r$ & $18.59 \pm 0.03$ \\ 
58244.170880 & 10.99 & P48 & $r$ & $18.47 \pm 0.02$ \\ 
58245.171447 & 12.00 & P48 & $r$ & $18.32 \pm 0.02$ \\ 
58245.172384 & 12.00 & P48 & $r$ & $18.31 \pm 0.02$ \\ 
58246.233762 & 13.06 & P48 & $r$ & $18.32 \pm 0.03$ \\ 
58247.234363 & 14.06 & P48 & $r$ & $18.28 \pm 0.02$ \\ 
58247.358800 & 14.18 & P60 & $$r$$ & $18.30 \pm 0.04$ \\ 
58248.235324 & 15.06 & P48 & $r$ & $18.21 \pm 0.02$ \\ 
58248.236250 & 15.06 & P48 & $r$ & $18.17 \pm 0.02$ \\ 
58248.335300 & 15.16 & P60 & $$r$$ & $18.21 \pm 0.03$ \\ 
58249.234444 & 16.06 & P48 & $r$ & $18.23 \pm 0.03$ \\ 
58250.234803 & 17.06 & P48 & $r$ & $18.17 \pm 0.03$ \\ 
58254.191401 & 21.02 & P48 & $r$ & $18.31 \pm 0.02$ \\ 
58254.192338 & 21.02 & P48 & $r$ & $18.26 \pm 0.02$ \\ 
58255.238356 & 22.06 & P48 & $r$ & $18.32 \pm 0.02$ \\ 
58256.217651 & 23.04 & P48 & $g$ & $19.02 \pm 0.05$ \\ 
58256.218113 & 23.04 & P48 & $g$ & $18.98 \pm 0.05$ \\ 
58256.218565 & 23.04 & P48 & $g$ & $19.06 \pm 0.06$ \\ 
58256.219028 & 23.04 & P48 & $g$ & $19.05 \pm 0.05$ \\ 
58256.219479 & 23.04 & P48 & $g$ & $19.05 \pm 0.04$ \\ 
58256.219942 & 23.04 & P48 & $g$ & $19.02 \pm 0.03$ \\ 
58256.220393 & 23.04 & P48 & $g$ & $19.03 \pm 0.02$ \\ 
58256.220845 & 23.04 & P48 & $g$ & $19.07 \pm 0.03$ \\ 
58256.221308 & 23.05 & P48 & $g$ & $19.11 \pm 0.02$ \\ 
58256.221759 & 23.05 & P48 & $g$ & $19.05 \pm 0.02$ \\ 
58256.222222 & 23.05 & P48 & $g$ & $19.04 \pm 0.03$ \\ 
58256.222674 & 23.05 & P48 & $g$ & $19.04 \pm 0.03$ \\ 
58256.223125 & 23.05 & P48 & $g$ & $19.04 \pm 0.03$ \\ 
58256.223588 & 23.05 & P48 & $g$ & $19.12 \pm 0.03$ \\ 
58256.244317 & 23.07 & P48 & $r$ & $18.40 \pm 0.03$ \\ 
58256.278032 & 23.10 & P48 & $r$ & $18.43 \pm 0.02$ \\ 
58257.232951 & 24.06 & P48 & $r$ & $18.42 \pm 0.03$ \\ 
58257.233877 & 24.06 & P48 & $r$ & $18.41 \pm 0.03$ \\ 
58258.168634 & 24.99 & P48 & $g$ & $19.32 \pm 0.04$ \\ 
58262.202593 & 29.03 & P48 & $r$ & $18.72 \pm 0.04$ \\ 
58262.220127 & 29.04 & P48 & $g$ & $19.53 \pm 0.08$ \\ 
58262.252870 & 29.08 & P48 & $r$ & $18.64 \pm 0.05$ \\ 
58263.235185 & 30.06 & P48 & $r$ & $18.74 \pm 0.03$ \\ 
58263.259248 & 30.08 & P48 & $g$ & $19.81 \pm 0.11$ \\ 
58266.250648 & 33.07 & P48 & $r$ & $19.02 \pm 0.07$ \\ 
58266.251562 & 33.08 & P48 & $r$ & $19.08 \pm 0.07$ \\ 
58267.185671 & 34.01 & P48 & $g$ & $20.08 \pm 0.29$ \\ 
58267.290174 & 34.11 & P48 & $r$ & $18.91 \pm 0.07$ \\ 
58268.167917 & 34.99 & P48 & $g$ & $20.13 \pm 0.24$ \\ 
58269.185035 & 36.01 & P48 & $r$ & $19.20 \pm 0.07$ \\ 
58269.185972 & 36.01 & P48 & $r$ & $19.08 \pm 0.06$ \\ 
58270.173681 & 37.00 & P48 & $r$ & $19.28 \pm 0.05$ \\ 
58272.184954 & 39.01 & P48 & $r$ & $19.37 \pm 0.04$ \\ 
58272.185880 & 39.01 & P48 & $r$ & $19.26 \pm 0.04$ \\ 
58274.198912 & 41.02 & P48 & $r$ & $19.45 \pm 0.05$ \\ 
58274.234641 & 41.06 & P48 & $g$ & $20.58 \pm 57.42$ \\ 
58276.198576 & 43.02 & P48 & $r$ & $19.67 \pm 0.05$ \\ 
58276.199502 & 43.02 & P48 & $r$ & $19.50 \pm 0.05$ \\ 
58276.213970 & 43.04 & P48 & $g$ & $20.51 \pm 0.11$ \\ 
58276.214907 & 43.04 & P48 & $g$ & $20.56 \pm 0.11$ \\ 
58277.193495 & 44.02 & P48 & $g$ & $20.75 \pm 0.15$ \\ 
58277.243113 & 44.07 & P48 & $r$ & $19.51 \pm 0.06$ \\ 
58278.194016 & 45.02 & P48 & $g$ & $20.32 \pm 0.13$ \\ 
58278.237199 & 45.06 & P48 & $r$ & $19.62 \pm 0.07$ \\ 
58279.171516 & 46.00 & P48 & $r$ & $19.63 \pm 0.08$ \\ 
58279.187500 & 46.01 & P48 & $r$ & $19.63 \pm 0.06$ \\ 
58279.207593 & 46.03 & P48 & $g$ & $20.63 \pm 0.13$ \\ 
58279.208530 & 46.03 & P48 & $g$ & $20.60 \pm 0.12$ \\ 
58280.174988 & 47.00 & P48 & $r$ & $19.63 \pm 0.09$ \\ 
58280.227755 & 47.05 & P48 & $g$ & $20.85 \pm 0.15$ \\ 
58281.194468 & 48.02 & P48 & $r$ & $19.76 \pm 0.07$ \\ 
58281.237141 & 48.06 & P48 & $g$ & $20.65 \pm 0.14$ \\ 
58282.193773 & 49.02 & P48 & $r$ & $19.77 \pm 0.07$ \\ 
58282.194699 & 49.02 & P48 & $r$ & $19.82 \pm 0.07$ \\ 
58282.243113 & 49.07 & P48 & $g$ & $20.51 \pm 0.14$ \\ 
58282.244039 & 49.07 & P48 & $g$ & $20.58 \pm 0.15$ \\ 
58283.215544 & 50.04 & P48 & $r$ & $19.81 \pm 0.07$ \\ 
58283.237836 & 50.06 & P48 & $g$ & $20.45 \pm 0.13$ \\ 
58284.203982 & 51.03 & P48 & $r$ & $19.80 \pm 0.08$ \\ 
58284.214236 & 51.04 & P48 & $g$ & $20.83 \pm 0.17$ \\ 
\enddata 
\tablenotetext{}{\textbf{Note.} Values have been corrected for Milky Way extinction. Phase is relative to $t_0$ defined in Section \ref{sec:discovery}.}
\end{deluxetable} 

\subsection{Spectral Classification}
\label{sec:spectra}

A log of our spectroscopic follow-up observations of \name\ is provided in Table \ref{tab:ZTF18aaqjovh-spectra}.

On 2018 May 9 UT we obtained a spectrum of \name\ using the SEDM and compared it to a set of spectral templates from the publicly available Supernova Identification code (SNID; \citealt{Blondin2007}).
The best match was to a spectrum of SN\,1998bw taken at 16 days post-explosion.
As shown in Figure \ref{fig:lc},
a comparison with the light curve of SN\,1998bw suggests that these two spectra were obtained at comparable phases.
So, we classified \name\ as Type~Ic-BL.

On 2018 May 14 UT, we observed \name\ using the Low Resolution Imaging Spectrometer \citep{Oke1995} on the Keck I 10m telescope.
The spectrum was reduced and extracted using \texttt{LPipe} \citep{Perley2019}.
The next day, we observed the source using
the Andalusia Faint Object Spectrograph and Camera (ALFOSC\footnote{http://www.not.iac.es/instruments/alfosc/}) on the Nordic Optical Telescope (NOT; \citealt{Djupvik2010}).
The NOT spectrum was reduced in a standard way, including wavelength calibration against an arc lamp, and flux calibration using a spectrophotometric standard star.
We obtained another spectrum on 2018 June 8 UT
using the Double Beam Spectrograph (DBSP; \citealt{Oke1982}) on the 200 inch Hale telescope at the Palomar Observatory.
The DBSP spectrum was reduced using a PyRAF-based pipeline
\citep{Bellm2016}.
We obtained a final spectrum one month later using LRIS.

\begin{figure}
\centering
\includegraphics[width=\columnwidth]{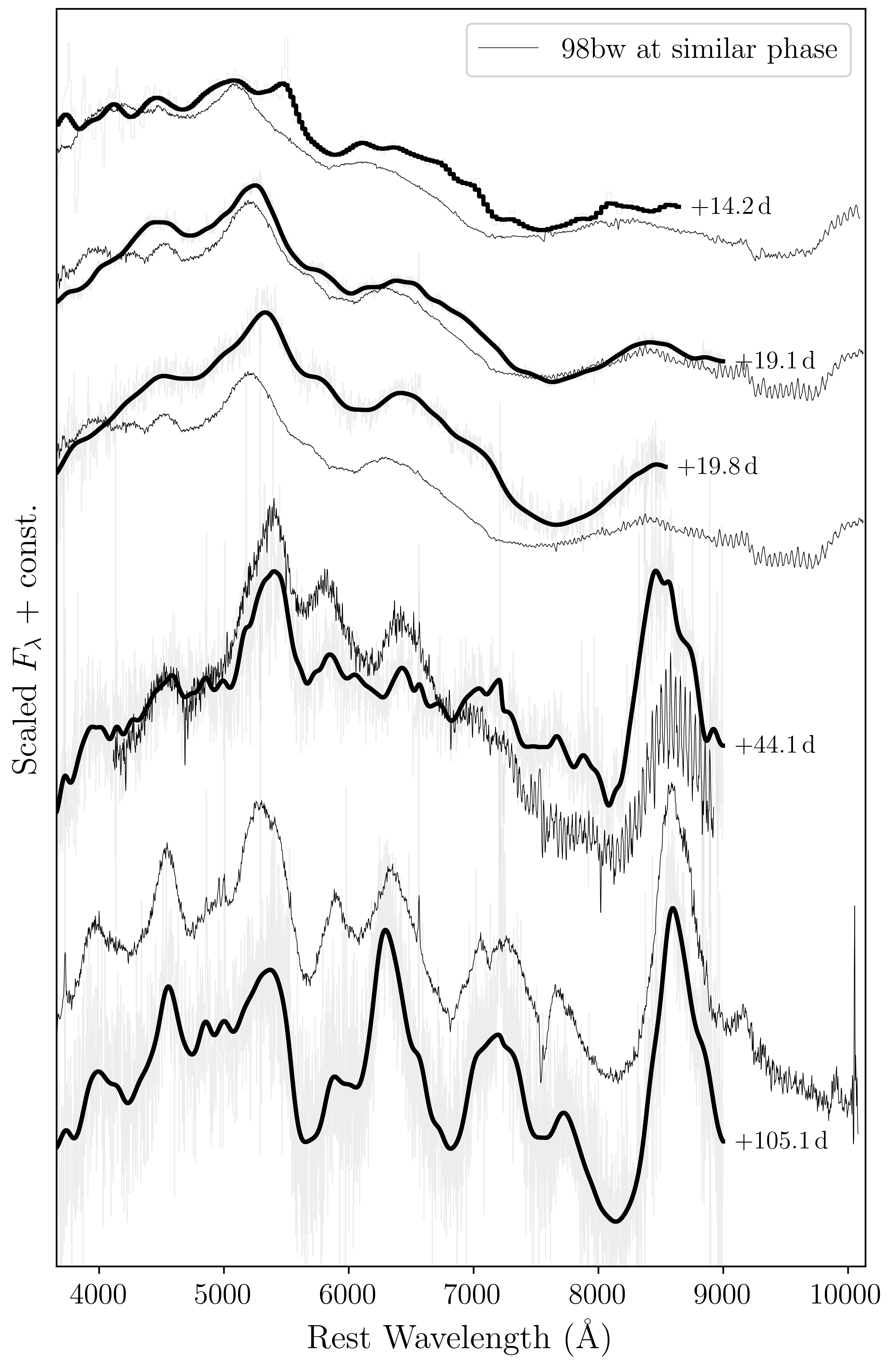}
\caption{Optical spectra of \name. Full spectra are shown in light gray and smoothed spectra are shown in thick black lines.
For comparison, we show spectra of SN\,1998bw at similar phases as thin black lines.
The SN\,1998bw spectra were taken from the Open Supernova Catalog (\href{https://sne.space/}{https://sne.space/})
and are originally from \citet{Patat2001}.
}
\label{fig:spec-sequence}
\end{figure}

\begin{figure}
\centering
\includegraphics[width=1.0\columnwidth]{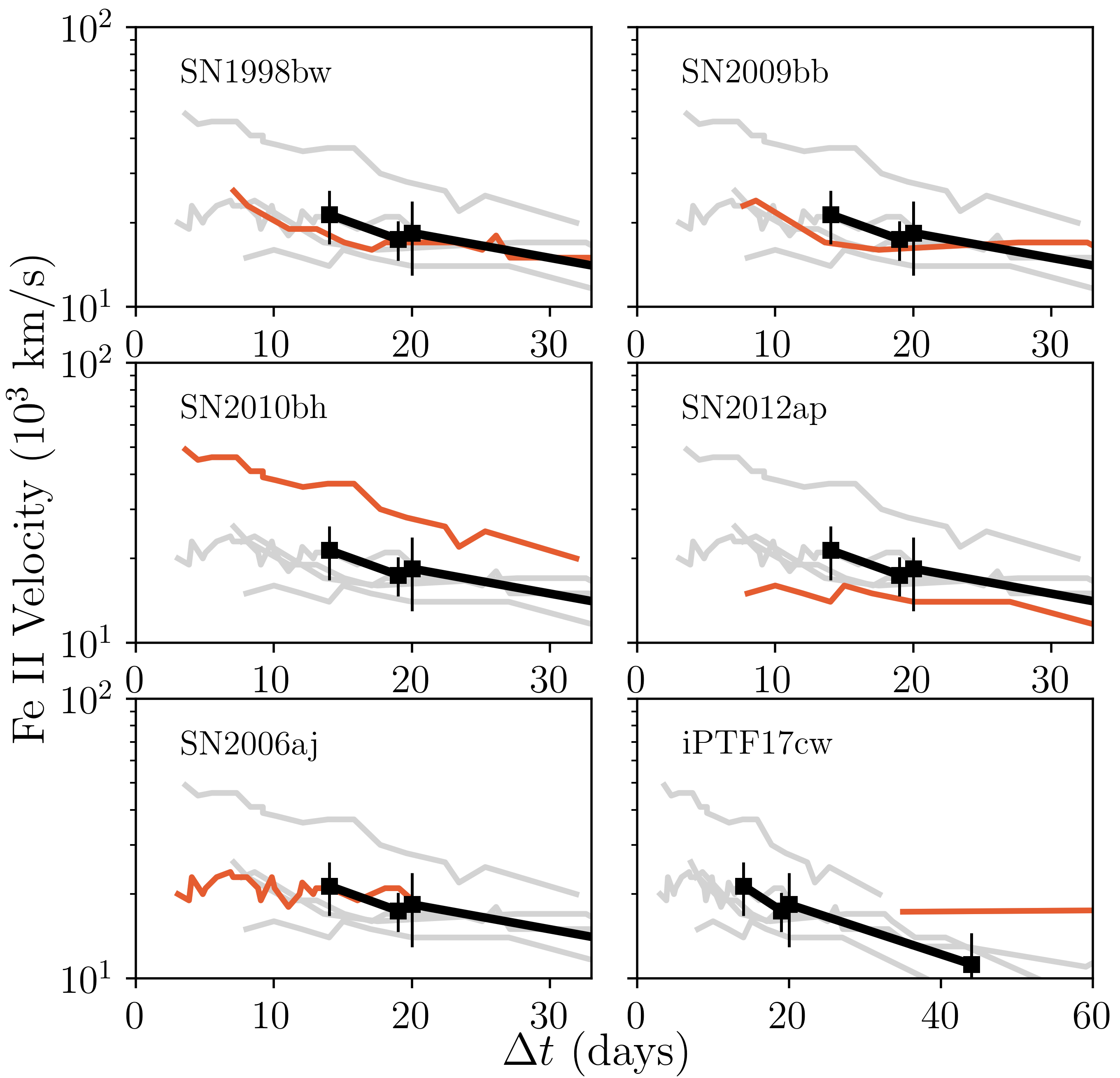}
\caption{Evolution of the photospheric velocity of \name\ over time as measured from Fe II absorption features in the Ic-BL spectra.
For comparison, we show the velocity evolution of several LLGRB-SNe (SN\,1998bw / GRB\,908425, SN\,2010bh / GRB\,100316D, and SN\,2006aj / GRB\,060218) and radio-loud relativistic SNe lacking a coincident GRB detection (SN\,2009bb, SN\,2012ap, and iPTF\,17cw).
Each panel shows measurements for \name\ as black squares, the population of comparison events as light gray lines in the background,
and one comparison SN highlighted in orange.
Data were taken from \citet{Modjaz2016} and explosion times were estimated from \citet{Galama1998}, \citet{Campana2006}, \citet{Soderberg2010}, \citet{Bufano2012}, \citet{Milisavljevic2015}, and \citet{Corsi2014}.
}
\label{fig:vel}
\end{figure}

The spectral sequence obtained via our follow-up for \name\ is shown in Figure \ref{fig:spec-sequence},
compared to spectra of SN\,1998bw at similar phases post-explosion.
We used our spectra to estimate the photospheric velocity of \name\ as a function of time.
In typical Ic SNe, photospheric velocity is measured using the width of the \ion{Fe}{2} $\lambda$5169 line (e.g., \citealt{Branch2002}).
However, due to the high velocities in Ic-BL SNe,
the \ion{Fe}{2} $\lambda$5169 line is blended with the nearby \ion{Fe}{2} $\lambda\lambda$4924,5018 lines.
So, to perform our velocity measurements,
we use the publicly available code\footnote{https://github.com/nyusngroup/SESNspectraLib} based on the method in \citet{Modjaz2016},
which convolves a Ic spectrum with Gaussian functions of varying widths until a best match is reached.
For the SEDM measurements, we subtracted the contribution to the velocity from the resolution of the spectrograph,
assuming that $\Delta v_\mathrm{obs}^2 = \Delta v_\mathrm{real}^2 + \Delta v_\mathrm{inst}^2$
and that $\Delta v_\mathrm{inst}=3000\,\km\,\psec$.
The resulting velocities are listed in Table \ref{tab:ZTF18aaqjovh-spectra}, and we show the velocity evolution compared to other Ic-BL SNe in Figure \ref{fig:vel}.

\begin{deluxetable}{lrrrr}[htb!]
\tablecaption{Spectroscopic observations of \name \label{tab:ZTF18aaqjovh-spectra}} 
\tablewidth{0pt} 
\tablehead{\colhead{Date} & \colhead{$\Delta t$} & \colhead{Tel.+Instr.} & \colhead{Exp. Time} & \colhead{$v_\mathrm{ph}$} \\ 
\colhead{(MJD)} & \colhead{(days)} & & \colhead{(s)} & ($10^{4}\,\km\,\psec$)} 
\tabletypesize{\scriptsize} 
\startdata 
58247.359 & 14 & P60+SEDM & 1800 & $2.12 \pm 0.46$ \\
58252.322 & 19 & Keck I+LRIS & 920 & $1.74 \pm 0.28$ \\
58253.977 & 20 & NOT+ALFOSC & 2400 & $1.84 \pm 0.54 $ \\
58277.253 & 44 & P200+DBSP & 2700 & $1.12 \pm 0.33 $ \\
58338.249 & 105 & Keck I+LRIS & 1720 & N/A \\
\enddata 
\end{deluxetable}

\subsection{Radio Observations}
\label{sec:radio-obs}

Upon classifying \name\ as a Type Ic-BL SN (Section \ref{sec:spectra})
we triggered the VLA for radio follow-up observations 
under the program VLA/18A-176 (PI: A. Corsi).
A log of our observations is provided in Table \ref{tab:radio-flux}.

We observed the field of \name\ with the VLA over several epochs using the S, C, and Ku bands. We used J1150+2417 as our complex gain calibrator, and 3C286 as our flux density and bandpass calibrator. Data were calibrated using the VLA calibration pipeline available in the Common Astronomy Software Applications (CASA; \citealt{McMullin2007}). After calibration, we inspected the data manually for further flagging. Images of the field were created using the CLEAN algorithm \citep{Hogbom1974} available in CASA.

In our VLA images, we found a radio point source consistent with the optical position of \name. Although the radio emission from this source remained fairly constant during the three epochs of our monitoring in C-band (see Table \ref{tab:radio-flux}), its transient nature was confirmed by a nondetection about $280$ days after the SN optical discovery. The radio peak flux densities are reported in Table \ref{tab:radio-flux}. Flux density errors are calculated as the quadrature sum of the image RMS and a fractional $5\%$ absolute flux calibration error.

\begin{figure*}
\centering
\includegraphics[width=0.6\textwidth]{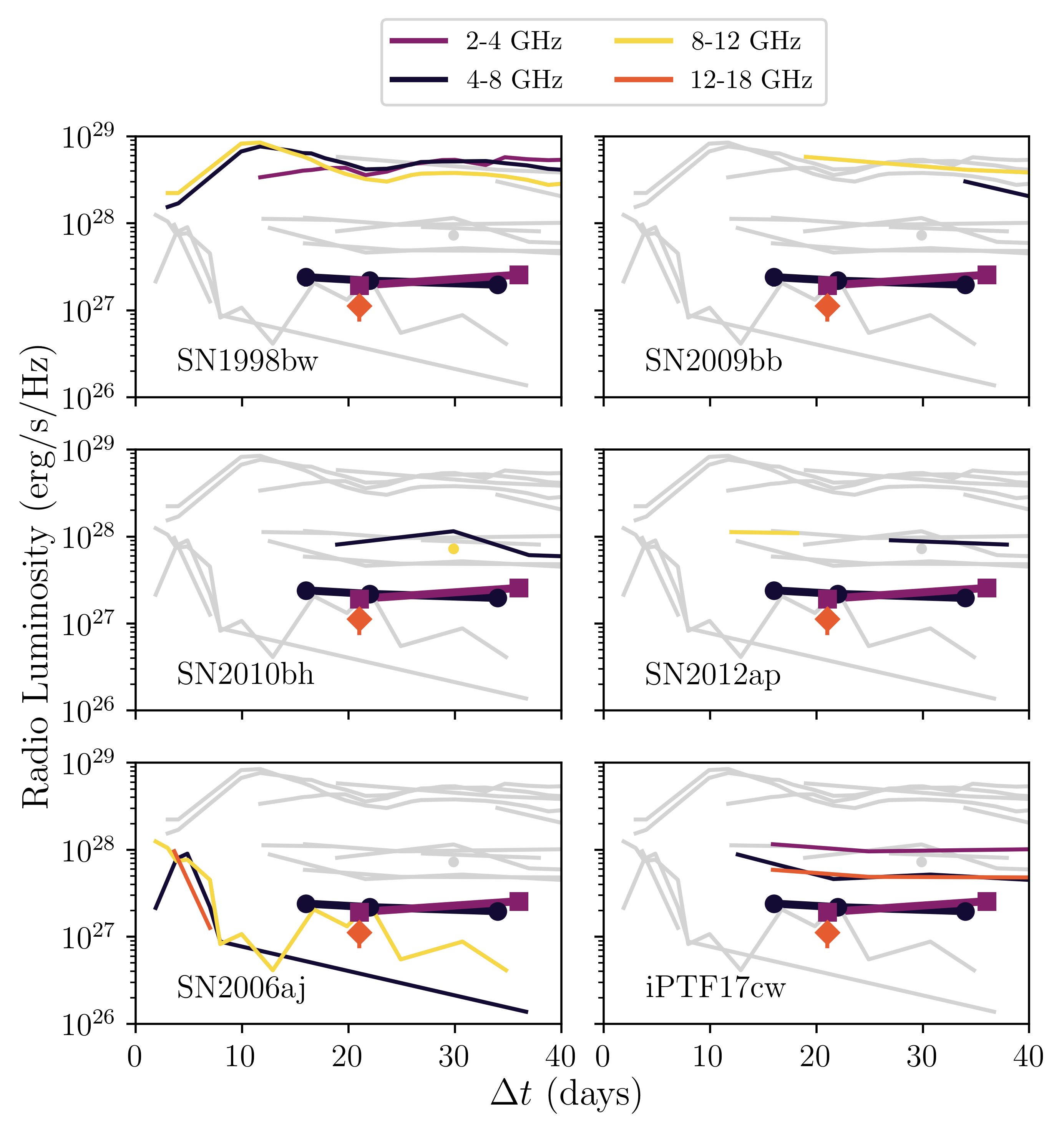}
\caption{
Radio light curve of \name\
compared with LLGRB-SNe (SN\,1998bw / GRB\,980425, SN\,2010bh / GRB\,100316D, and SN\,2006aj / GRB\,060218) and relativistic SNe (SN\,2009bb, SN\,2012ap, and iPTF17cw).
Each panel shows observations of \name\ (connected symbols), the population of comparison events as light gray lines in the background,
and one SN highlighted as colored lines for comparison.
Note that \name\ lacks data in the 8--12\,\ghz\ range.
Data were taken from \citet{Kulkarni1998}, \citet{Soderberg2010}, \citet{Chakraborti2015}, \citet{Margutti2014}, \citet{Soderberg2006}, and \citet{Corsi2017}.
}
\label{fig:radio-lum}
\end{figure*}

The radio light curve of \name\ is shown in Figure \ref{fig:radio-lum}, compared to several other Ic-BL SNe. 
At the distance of \name, the 6\,\ghz\ radio luminosity density at $\Delta t \approx 20\,\days$ since explosion is $2 \times 10^{27}\,\erg\,\psec\,\phz$. This is over an order of magnitude fainter than SN\,1998bw at a similar epoch, and most similar to the luminosity of iPTF17cw at similar frequencies.

\begin{deluxetable*}{lccccccc}[!htb]
\tablecaption{Radio flux density measurements of \name\label{tab:radio-flux}} 
\tablewidth{0pt} 
\tablehead{ 
\colhead{Start Date} & \colhead{Time on-source} & \colhead{$\Delta t$} &\colhead{$S_{3\,\rm GHz}$  } & \colhead{$S_{6\,\rm GHz}$ } & \colhead{$S_{15\,\rm GHz}$} & \colhead{Array Config.} \\ 
\colhead{(UT)} & \colhead{(hr)} & \colhead{(days)} &\colhead{$(\mu$Jy) } & \colhead{$(\mu$Jy) } & \colhead{$(\mu$Jy) } & 
} 
\tabletypesize{\scriptsize} 
\startdata 
2018 May 11 & 0.67 & 16 &-- & $32.5 \pm 7.1$ & --&  A   \\
2018 May 16 & 0.67 & 21 &$26.0 \pm 6.9$ & -- &$15.1 \pm 5.2$ &  A  \\
2018 May 17 & 0.67 & 22 &-- & $29.6 \pm 5.3$ & --  & A  \\
2018 May 29 & 0.67 & 34 & --& $26.6 \pm 5.4$ &-- &  A \\
2018 May 31 & 1.5 & 36 & $34.6 \pm 4.8$ & --& --& A \\
2019 Jan 26 & 1.5 & 276 &-- & $\lesssim 15$& --&  C 
\enddata 
\end{deluxetable*}

\subsection{X-Ray Observations}
\label{sec:xray-obs}

A log of our X-ray observations is provided in Table \ref{tab:ZTF18aaqjovh-xray}.

On 2018 May 31 UT,
we obtained a 2.5\,ks target-of-opportunity observation of the position of \name\ with the X-ray Telescope (XRT) on board the Neil Gehrels Swift Observatory \citep{Burrows2005}.
We built the XRT light curve using the online generator \citep{Evans2009}.
On the web form\footnote{https://www.swift.ac.uk/user\_objects/}, we used the default values except for \texttt{Try to centroid?}, which was set to \texttt{No}.
The source was not detected with a 3$\sigma$ upper limit of $7.2 \times 10^{-3}\,$cps.
To convert the upper limit from count rate to flux, we assumed a Galactic neutral hydrogen column density\footnote{https://heasarc.gsfc.nasa.gov/cgi-bin/Tools/w3nh/w3nh.pl} of $n_H = 1.37 \times 10^{20}\,\pcmsq$ and a power-law spectrum $f \propto E^{-\Gamma}$ where $f$ is flux (photons\,\pcmsq\,\psec), $E$ is energy, and $\Gamma=2$ is the photon index. This gives an unabsorbed upper-limit on the 0.3--10\,\kev\ flux of $2.3 \times 10^{-13}\,\erg\,\pcmsq\,\psec$, corresponding to a luminosity of $1.7 \times 10^{42}\,\erg\,\psec$.

\begin{figure}[!htb]
\centering
\includegraphics[width=1.0\columnwidth]{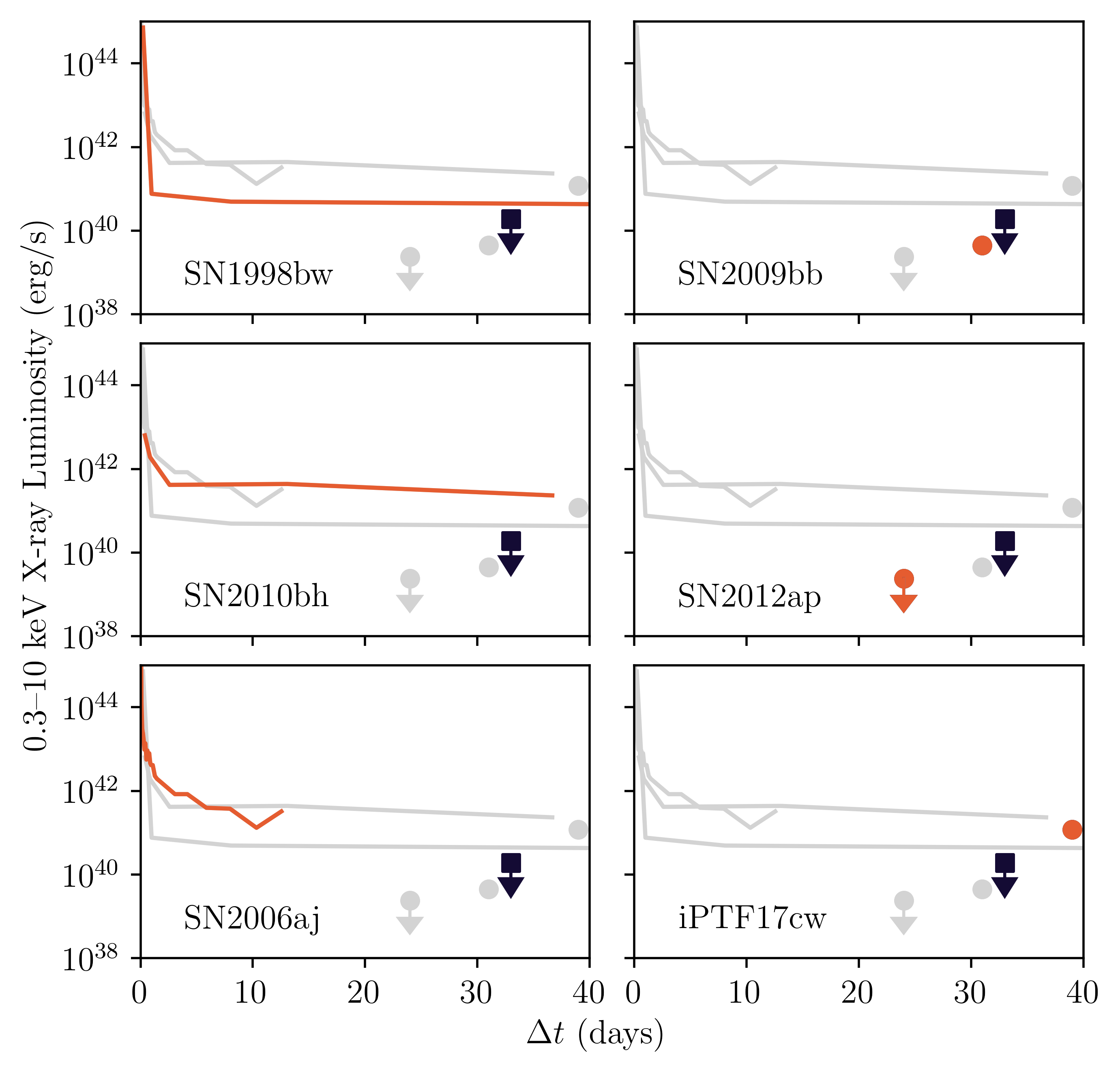}
\caption{Upper limit on the X-ray luminosity of ZTF18aaqjovh from our first Chandra observation (black square) compared to the X-ray luminosity at similar epochs of LLGRBs (SN\,1998bw, SN\,2010bh, and SN\,2006aj) and Ic-BL SNe with relativistic outflows discovered independently of a $\gamma$-ray trigger (iPTF17cw, SN\,2009bb, and SN\,2012ap).
Each panel shows the full set of comparison events in light gray, with one event highlighted in orange.
Data were taken from \citet{Corsi2017}, \citet{Campana2006}, \citet{Margutti2014}.
}
\label{fig:xray-lum}
\end{figure}

We also obtained two epochs of observations of  \name\ with the Advanced CCD Imaging Spectrometer (ACIS; \citealt{Garmire2003}) on the Chandra X-ray Observatory via our approved program (No. 19500451, PI: Corsi). The first epoch began at 11:07 on 2018 May 28 UT ($\Delta t \approx 33\,\days$) under ObsId 20315 (integration time 9.93 ks), and the second began at 11:10 on 2018 July 24 UT ($\Delta t \approx 90\,\days$) under ObsId 20316.
No X-ray emission was detected at the location of \name\ in either epoch, with a 90\% upper limit on the 0.5--7.0\,\kev\ count rate of $2.52\times10^{-4}$\,\counts\,\psec\ and $2.32\times10^{-4}$\,\counts\,\psec, respectively.
For the same Galactic $n_H$ and power-law source model that we used in the Swift data,
we obtain an upper limit on the unabsorbed 0.3--10\,keV flux of $3.4 \times 10^{-15}\,\erg\,\psec\,\pcmsq$ in the first epoch
and $3.1 \times 10^{-15}\,\erg\,\psec\,\pcmsq$ in the second epoch.
At the distance of \name, these correspond to upper limits on the X-ray luminosity of $2.5 \times 10^{40}\,\erg\,\psec$ and $2.3 \times 10^{40}\,\erg\,\psec$. These upper limits are compared with the X-ray luminosity of radio-loud Ic-BL SNe in Figure \ref{fig:xray-lum}.

\begin{deluxetable}{lrrrr}[!htb]
\tablecaption{X-Ray Observations of \name \label{tab:ZTF18aaqjovh-xray}} 
\tablewidth{0pt} 
\tablehead{ 
\colhead{Start Date} & 
\colhead{$\Delta t$} &
\colhead{Instr.} & 
\colhead{Int.} 
& \colhead{Flux} \\
\colhead{(UT)} & 
\colhead{(days)} &
\colhead{} & 
\colhead{(ks)} 
& \colhead{(\erg\,\psec\,\pcmsq)}} 
\tabletypesize{\scriptsize} 
\startdata 
2018 May 28 11:07:06 & 33 & Chandra/ACIS & 9.93 & $<3.4 \times 10^{-15}$ \\
2018 May 31 00:33:57 & 36 & Swift/XRT & 2.5 & $<2.3 \times 10^{-13}$ \\
2018 July 24 11:10:42 & 90 & Chandra/ACIS & 9.93 & $<3.1 \times 10^{-15}$ \\
\enddata 
\end{deluxetable}

\subsection{Search for Gamma-Rays}
\label{sec:high-energy-lims}

We searched for any gamma-ray burst (GRB) coincident with the position and estimated time of first light of ZTF18aaqjovh.
As shown in Figure \ref{fig:lc} and discussed in more detail in Section \ref{sec:opt-lc-modeling},
we can use the relative time between GRB 980425 and the $r$-band peak of SN\,1998bw to estimate the time of a GRB associated with ZTF18aaqjovh.
If this relative time is the same between the two SNe,
then the associated GRB would have been approximately at the time
of the last nondetection ($t_0\approx$ 2018 April 25 UT), 10 days prior to the first detection on 2018 May 5 UT.

To be conservative, we set our search window to be $t_0 \pm 10$\,\days.
In Table \ref{tab:grbs} we list all 20 GRBs detected in this window.
Of the 20, all but one are ruled out based on the position of the SN.
The only possible counterpart is a GRB on 2018 May 3 03:41:01 ($\Delta t=8$) detected by Konus-Wind while Fermi/GBM was offline. 
The duration of this burst was 35\,s. Modeling the spectrum with a cutoff power-law model with $E_p=107^{+64}_{-25}$\,\kev\ and 20--1500\,\kev\ fluence $2 \times 10^{-6}\,\erg\,\pcmsq$ we obtain an $L_\mathrm{iso}=8\times10^{47}\,\erg\,\psec$,
which is typical of LLGRBs \citep{Cano2017}.

\begin{deluxetable}{lrrrrr}[!htb]
\tablecaption{Gamma-Ray Bursts within 10 Days of the Estimated Time of First Light of ZTF18aaqjovh \label{tab:grbs}} 
\tablewidth{0pt} 
\tablehead{ 
\colhead{Date} & 
\colhead{Name} & 
\colhead{$\Delta t$} &
\colhead{Instr.} & 
\colhead{Pos.} &
\colhead{Verdict}
\\
\colhead{(UT)} & 
\colhead{} &
\colhead{(days)} &
\colhead{} & 
\colhead{} &
} 
\tabletypesize{\scriptsize} 
\startdata
2018 Apr 16 & 180416D & -9 & KAI & N               & N(a) \\
2018 Apr 16 & 180416A & -9 & KGI & 113.65, +49.120 & N(b) \\
2018 Apr 16 & 180416B & -9 & KGAC& 354.233, +78.433& N(b) \\
2018 Apr 17 &         & -8 & K   & S               & N(c) \\
2018 Apr 20 &         & -5 & KG  & 93.510, -28.320 & N(b) \\
2018 Apr 20 &         & -5 & KGI & 83.230, -25.250 & N(b) \\
2018 Apr 21 &         & -4 & K   & N               & N(c) \\
2018 Apr 23 &         & -2 & KGI & 208.680, +9.840 & N(b) \\
2018 Apr 25 & 180425A & 0  & KS  & 64.452, -32.952 & N(b) \\
2018 Apr 26 &         & 1  & KGI & 251.240, +81.390& N(b)\\
2018 Apr 26 &         & 1  & KG  & 202.410, +58.170& N(b) \\
2018 Apr 26 &         & 1  & K   & N               & N(c) \\
2018 Apr 26 &         & 1  & K   & S               & N(b) \\
2018 Apr 27 & 180427A & 2  & KGI & 283.330, +70.300& N(b) \\
2018 Apr 28 &         & 3  & KGI & 92.120, +54.780 & N(b) \\
2018 Apr 28 &         & 3  & K   & N               & N(c) \\
2018 Apr 29 &         & 4  & KI  & S               & N(b) \\
2018 May 3 &         & 8  & K   & N               & Y \\
2018 May 4 &         & 9  & KGI & 220.230, +38.720& N(b) \\
2018 May 4 & 180504A & 9  & KSI & 331.144, -14.658& N(b) \\
\enddata 
\tablecomments{In the Position column, N and S mean that the position is localized to the northern and southern ecliptic hemispheres, respectively. In the Instrument column, K means Konus-Wind, A means Astrosat, I means INTEGRAL SPI-ACS, G means Fermi/GBM, S means Swift/BAT. In the Verdict column, N means that an association is ruled out because (a) the SN position was Earth-occulted for Astrosat and GBM, (b) the SN position is inconsistent with the localized burst position, or (c) the SN position was visible to GBM but not detected. Y means that an association is possible.}
\end{deluxetable}

However, due to the coarse localization and the implication that the light curve of
ZTF18aaqjovh increased to peak brightness much more steeply than the light curve of SN\,1998bw,
we consider the association with the GRB on May 3 to be unlikely.
Assuming it is not related,
we can set a limit on the fluence and corresponding isotropic equivalent energy of a prompt burst associated with \name. 
The Interplanetary Network (IPN) has essentially a 100\% duty cycle across the sky,
and detects GRBs with $E_\mathrm{pk}>20\,\kev$ down to $6 \times 10^{-7}\,\erg\,\pcmsq$ at 50\% efficiency \citep{Hurley2010,Hurley2016}.
Using Konus-Wind waiting mode data near $t_0$ and assuming a typical GRB spectrum (a Band function with $\alpha=-1, \beta=-2.5$, and $E_p=300\,\kev$; \citealt{Band1993,Preece2000}), we estimate a peak limiting flux of $2.1\times10^{-7}\,\erg\,\pcmsq\,\psec$ (20--1500\,\kev, 2.944\,s scale).
At the distance of ZTF18aaqjovh, this corresponds to
an upper limit on a GRB peak luminosity of $L_\mathrm{iso} \approx 1.6 \times 10^{48}\,\erg\,\psec$,
two orders of magnitude less luminous than classical GRBs but similar to LLGRBs \citep{Cano2017}.
We note that the IPN would not be sensitive to LLGRBs such as LLGRB\,060218 associated with SN\,2006aj \citep{Cano2017} because of their soft spectra ($E_\mathrm{pk} < 20\,\kev$ for 060218).

\section{Analysis and Discussion}

\subsection{Modeling the Optical Light Curve}
\label{sec:opt-lc-modeling}

As shown in Figure \ref{fig:lc},
the $r$-band light curve of ZTF18aaqjovh declines slightly faster than the light curve of SN\,1998bw, and is 0.4\,mag fainter.
For an SN with an optical light curve powered by radioactive decay,
the ``stretch'' (width) of the light curve
scales with the kinetic energy and ejecta mass as \citep{Valenti2008,Lyman2016}

\begin{equation}
    \tau_m \propto \left(\frac{\mej^3}{E_k}\right)^{1/4},
\end{equation}

\noindent where $\tau_m$ is the width of the light curve,
\mej\ is the ejecta mass, and $E_k$ is the kinetic energy of the explosion.
The degeneracy between \mej\ and $E_k$ is broken by the photospheric velocity \citep[see Eq. 2 in][]{Lyman2016}:

\begin{equation}
    v_\mathrm{ph}^2 = \frac{5}{3} \frac{2E_k}{\mej}.
\end{equation}

As shown in Figure \ref{fig:lc} and Figure \ref{fig:vel},
ZTF18aaqjovh has a photospheric velocity close to that of SN\,1998bw, and its light curve is narrower.
So, we expect the ejecta mass and kinetic energy of \name\ to be slightly smaller to that of SN\,1998bw, which had $\mej \approx 4.4^{+1.2}_{-0.8}\,\msol$ and $E_k \approx 9.9^{+3.8}_{-2.2} \times 10^{51} \,\erg$, respectively \citep{Lyman2016},
values typical of Ic-BL SNe from untargeted surveys \citep{Taddia2019}.

Finally, assuming that the dominant powering mechanism for the optical light curve is radioactive decay, we have the following energy deposition rate from \nickel\ \citep{Kasen2017}:

\begin{multline}
    L_\mathrm{{}^{56} Ni}(t) = 2 \times 10^{43}
    \left( \frac{\mni}{\msol} \right)
    \bigl[ 3.9e^{-t/\tau_\mathrm{Ni}} \\ + 0.678  \left(e^{-t/\tau_\mathrm{Co}} - e^{-t/\tau_\mathrm{Ni}}\right) \bigr] \erg\,\psec
\end{multline}

\noindent where the decay lifetimes of \nickel\ and \cobalt\ are $\tau_\mathrm{Ni}=8.8\,\days$ and $\tau_\mathrm{Co}=113.6\,\days$, respectively.
Arnett's law \citep{Arnett1982} states that the instantaneous energy deposition rate is equal to the SN luminosity at peak.
Under this assumption, the peak luminosity is simply equal to $L_\mathrm{{}^{56} Ni}$ at that time,
so is directly proportional to \mni.

Taking $L\approx \nu L_\nu \approx 6.9 \times 10^{42}\,\erg\,\psec$ at peak light ($t\approx15\,\days$)
we find that $\mni \approx0.3\,\msol$.
For reference, the nickel mass of SN\,1998bw has been estimated to be $\mni \approx 0.54^{+0.08}_{-0.07}\,\msol$ \citep{Lyman2016}.
These values are typical for GRB-SNe \citep{Cano2017}
and for Ic-BL SNe in general \citep{Taddia2019}.

\subsection{Properties of the Fastest (Radio-emitting) Ejecta}

As shown in Figure \ref{fig:radio-lum},
the radio luminosity of ZTF18aaqjovh is between that of SN\,2006aj and that of iPTF17cw.
Due to the faintness of the SN it is unfortunately difficult to measure the true rate of change of the flux, but the slow temporal evolution of the 3-6\,GHz flux during the first four epochs of observation ($\Delta t=16\,\days$ to $\Delta t=36\,\days$) may imply that the synchrotron self-absorption (SSA) frequency is passing through these frequencies at this time.
This is supported by the 3--15\,\ghz\ observations at $\Delta t=21-22\,\days$,
which suggest that the SSA peak is below 15\,\ghz\ and close to 3--6\,\ghz.
Altogether, we conclude that the SSA peak is 3--15\,\ghz\ at $\Delta t\approx20\,\days$, and that the peak flux is 20--$30\,\mu$Jy.

With these estimates of the SSA peak frequency and peak flux,
we use the framework laid out in \citet{Chevalier1998} to estimate the
shock energy $U$ (the amount that has been converted into pressure by the ambient medium),
the ambient density,
and the mean shock velocity at $\Delta t\sim20\,\days$.
The assumption is that the synchrotron spectrum arises from a population of relativistic electrons with a power-law number distribution in Lorentz factor $\gamma_e$ and some minimum Lorentz factor $\gamma_m$:

\begin{equation}
    \frac{dN(\gamma_e)}{d\gamma_e} \propto \gamma_e^{-p}, \gamma_e \geq \gamma_m.
\end{equation}

For typical radio SNe, $2.5 < p < 3$ \citep{Jones1991}.
Here we assume $p\approx3$,
as in \citet{Chevalier1998}.
Under this assumption, expressions for the shock radius and magnetic field strength
are given in Equations (13) and (14) of \citet{Chevalier1998}.
The magnetic field strength can then be used to estimate the magnetic energy density, assuming that equal amounts of energy are partitioned into electrons, magnetic fields, and protons \citep{Soderberg2010}.

\begin{figure}[!ht]
\centering
\includegraphics[width=1.0\columnwidth]{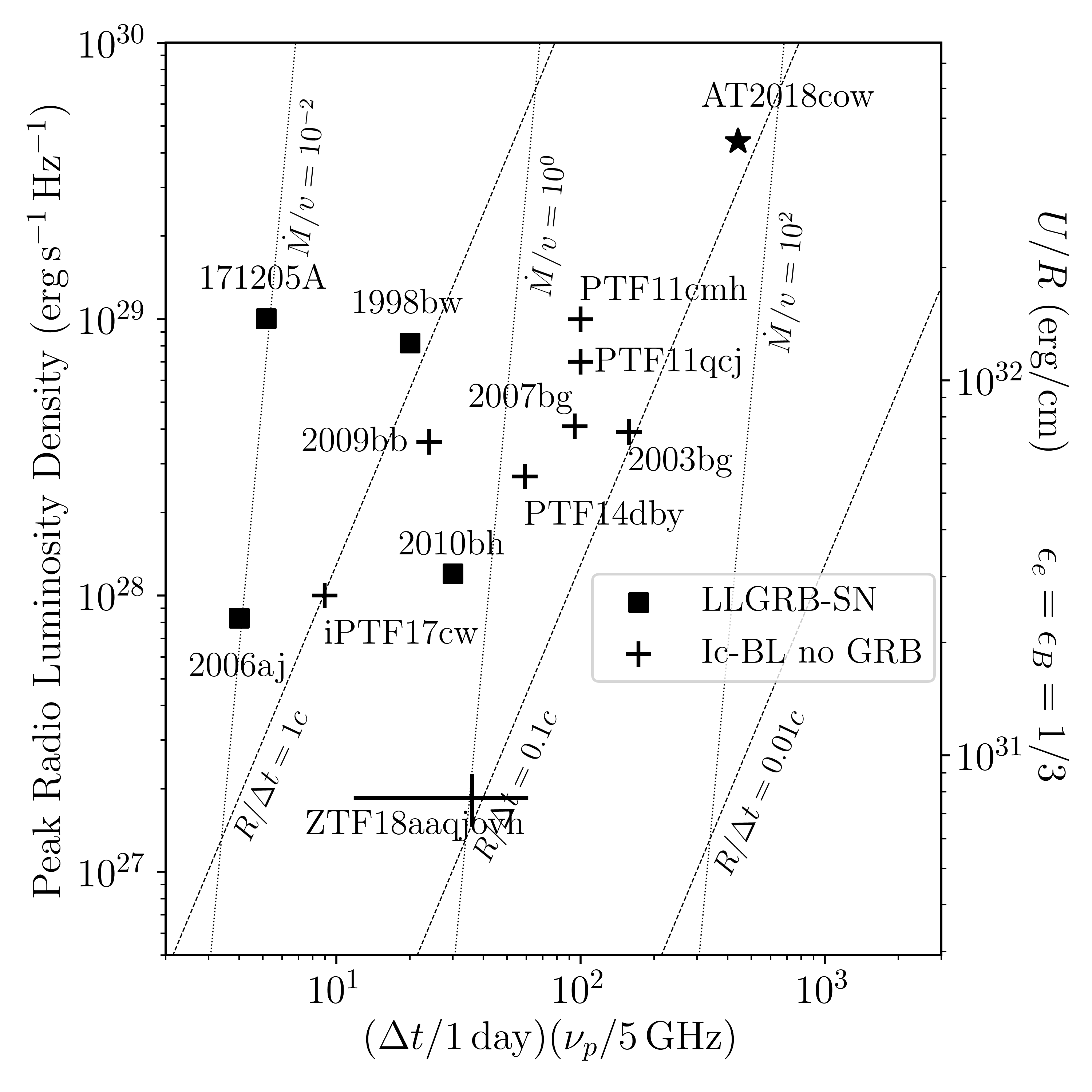}
\caption{Peak radio luminosity of ZTF18aaqjovh compared to other energetic stellar explosions; see \citet{Chevalier1998}, \citet{Soderberg2010}, and \citet{Ho2019}.
In \citet{Ho2019} we showed that the peak luminosity is directly proportional to $U/R$, the energy swept up per unit radius;
we display this value on the right-hand side.
Error bars reflect the estimated SSA peak (20--30\,$\mu$Jy, 3--15\,\ghz) at $\Delta t\approx20\,\days$.
Lines of constant velocity are shown, as well as lines of constant mass-loss rate (scaled to wind velocity) in units of $10^{-4}\,\msol \pyr /1000\,\km\,\psec$.
The radio luminosity for GRB 171205A was taken from VLA
observations  reported  by \citet{Laskar2017} but  we  note  that  this  is  a  lower  limit  in luminosity  and  in  peak  frequency  because  the  source  was  heavily  self-absorbed  at  this
epoch.
The radio luminosity for other sources is from, or derived using data from, \citet{Soderberg2010}, \citet{Kulkarni1998}, \citet{Soderberg2006}, \citet{Margutti2013}, \citet{Corsi2014}, \citet{Corsi2017}, \citet{Salas2013}, \citet{Soderberg2005}, \citet{Soderberg2006}.
}
\label{fig:radio-lum-tnu}
\end{figure}

These relations between observables and physical properties are summarized in Figure \ref{fig:radio-lum-tnu}, adapted from \citet{Ho2019}.
The mean velocity of the shock we derive for ZTF18aaqjovh is $v=0.06$--0.4$c$.
So, the outflow associated with ZTF18aaqjovh
could have been as fast as that observed in
the GRB-associated SN\,2010bh.
The implied mass-loss rate is
0.1--3 $\times 10^{-4} (v_w/1000\,\km\,\psec) \,\msol\,\pyr$,
which could be as high as that of the
strongly CSM-interacting SN PTF\,11qcj \citep{Corsi2014}.

\subsection{Modeling the Radio to X-Ray SED}

In SN explosions, the shockwave that accelerates electrons into a power-law distribution and produces synchrotron radiation,
detected as radio emission,
can also produce X-rays \citep{Chevalier2006}
via several mechanisms.
X-rays can have the same origin as the radio emission (lying along the same synchrotron spectrum). However, X-rays can also arise from inverse Compton scattering of the optical photons by the electrons producing the radio emission \citep{Chevalier2006,Chevalier2017}.
For a number of Ic-BL SNe, it seems that the simple synchrotron scenario is insufficient to explain the radio and X-ray observations --- in other words, there is an excess of X-ray emission \citep{Soderberg2006,Margutti2013,Corsi2014}.

As described in Section \ref{sec:xray-obs},
we do not detect X-ray emission from ZTF18aaqjovh,
corresponding to upper limits of $L_X < 3.4 \times 10^{40}\,\erg\,\psec$ at $\Delta t\sim33\,\days$ and $L_X < 3.1 \times 10^{40}\,\erg\,\psec$ at $\Delta t\sim90\,\days$.
At $\Delta t\sim33\,\days$,
this is smaller than the luminosity of X-ray emission associated with iPTF17cw, SN\,1998bw (GRB\,980425), and SN\,2010bh (GRB\,031203) at a similar epoch.
The 0.3--10\,\kev\ luminosity of SN\,2010bh at $\Delta t = 38\,\days$ was $2.4 \times 10^{41}\,\erg\,\psec$ \citep{Margutti2014},
which was already the least X-ray luminous LLGRB at this phase (second only to GRB\,980425).
Due to a lack of data later than 10\,\days\ we cannot rule out a luminosity similar to SN\,2006aj, SN\,2009bb, and SN\,2012ap \citep{Margutti2014}.

\begin{figure}[!ht]
\centering
\includegraphics[width=1.0\columnwidth]{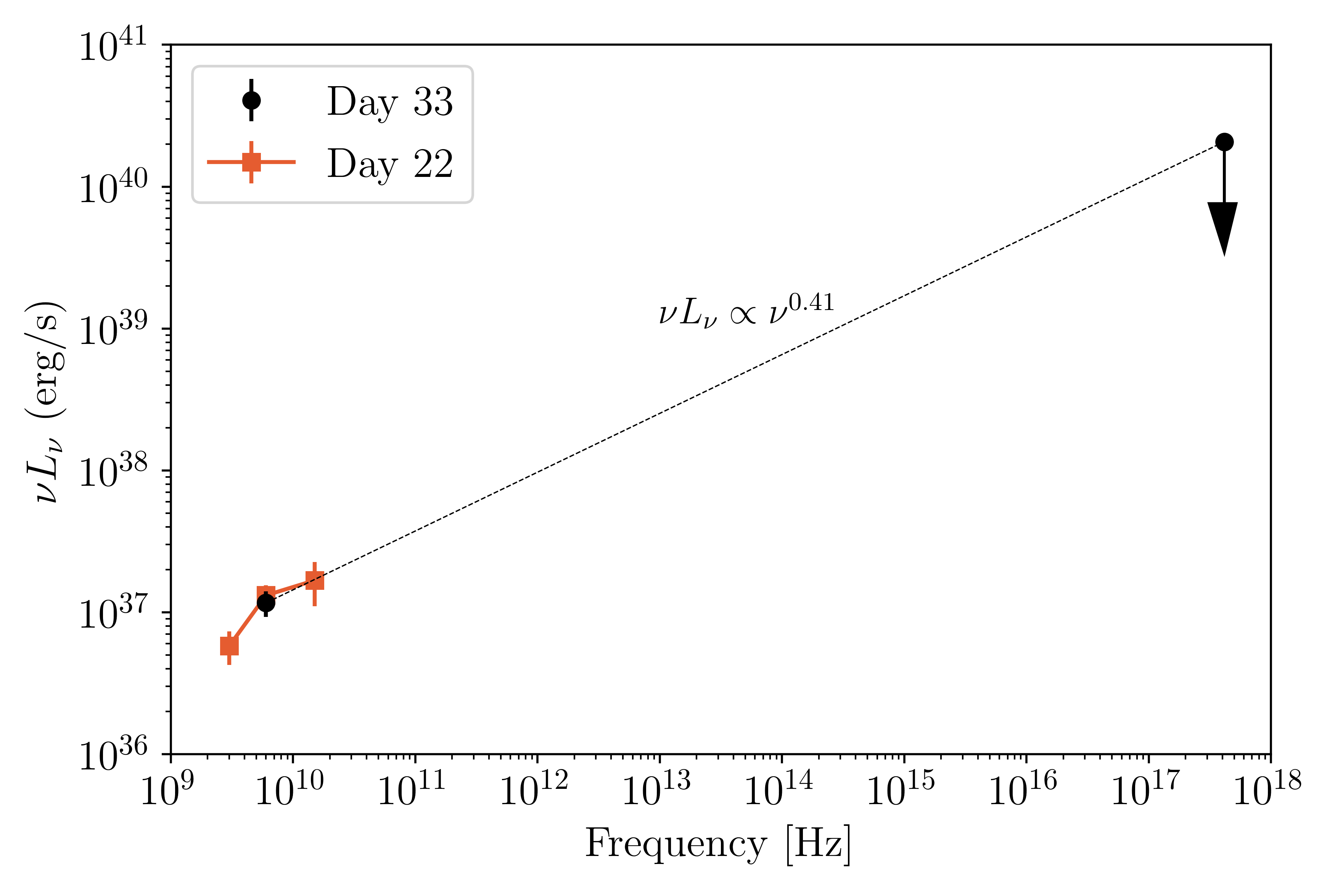}
\caption{Radio luminosity and upper limit on X-ray luminosity of ZTF18aaqjovh at $\Delta t\approx 33\,\days$.
From these measurements, we constrain the spectral index from the radio to X-ray frequencies to be $\beta<-0.6$ where
$L_\nu \propto \nu^{\beta}$.}
\label{fig:sed}
\end{figure}

Figure \ref{fig:sed} shows the radio luminosity and X-ray upper limit at $\Delta t \approx 33\,$days, from our observations of ZTF18aaqjovh with the VLA and Chandra on 2018 May 28--29 UT.
The spectral index is constrained to be $\beta<-0.6$ where $L_\nu \propto \nu^{\beta}$.
A common optically thin spectral index for radio SNe is $\beta\sim-0.5$ to $-1$ \citep{Chevalier1998} where $F_\nu \propto \nu^{\beta}$.
Above the cooling frequency,
this steepens to $\beta\sim -1$ or $\beta\sim-1.5$.
Thus we cannot conclude whether there is X-ray emission from an extension of the synchrotron spectrum,
or whether there is an excess from some other mechanism such as cosmic-ray-dominated shocks \citep{Chevalier2006}, which has been observed in a number of engine-driven SNe
including iPTF17cw ($\beta=-0.6$; \citealt{Corsi2017}), GRB\,060218 ($\beta=-0.5$; \citealt{Soderberg2006}), and GRB\,100316D ($\beta<-0.6$; \citealt{Margutti2014}).

\subsection{Gamma-Ray Burst}

In Section \ref{sec:high-energy-lims} we searched for coincident GRBs and found one possible counterpart,
although the association is highly unlikely due to the close proximity of the burst time with the first detection of the light curve.

Here we work under the hypothesis that ZTF18aaqjovh was associated with a GRB that we missed, and attempt to derive possible constraints on the $\gamma$-ray emission based on the SN properties.
From four GRB-SNe, \citet{Li2006} found the following relation between the peak spectral energy of the GRB and the peak bolometric luminosity of the associated SN:

\begin{equation}
    E_\mathrm{\gamma,peak} = 90.2\,\kev\
    \left( \frac{L_\mathrm{SN,peak}}{10^{43}\,\erg\,\psec} \right)^{4.97}.
\end{equation}

From the peak of the $r$-band light curve of \name,
we can estimate $L_\mathrm{SN,peak} \approx \nu f_\nu \approx 1.7 \times 10^{9}\,L_\odot$,
which gives $E_\mathrm{\gamma,peak} \approx 15$\,\kev. Using the so-called Amati relationship between a GRB peak energy and its isotropic equivalent energy \citep{Amati2006,Li2006}:

\begin{equation}
    E_\mathrm{\gamma,peak} = 97\,\kev\ \left(
    \frac{E_\mathrm{\gamma,iso}}{10^{52}\,\erg\,\psec} \right)^{0.49},
\end{equation}

\noindent we find an expected $E_\mathrm{iso} \approx 2 \times 10^{50}\,\erg$ for a potential GRB associated with \name.
These values of $E_\mathrm{\gamma,peak}$ and $E_\mathrm{iso}$ are similar to what has been measured for LLGRBs \citep{Cano2017},
and would not have been detectable by the IPN.

\section{Summary and Conclusions}

We have presented optical, X-ray, and radio observations of the Ic-BL SN ZTF18aaqjovh, discovered by ZTF as part of our campaign with the VLA to search for engine-driven explosions.
ZTF18aaqjovh shares a number of features in common with relativistic SNe:
an optical light curve similar to SN\,1998bw and
early peaking radio emission similar to iPTF17cw.
The limits on X-ray and gamma-ray emission rule out a classical GRB but cannot rule out an LLGRB.
Due to the low signal-to-noise of our measurements,
we can only constrain the velocity of the forward shock to be 0.06--0.4$c$.
Thus, this is at most a mildly relativistic explosion,
and we have no definitive evidence of a long-lived central engine.

From radio follow-up observations of Ic-BL SNe discovered by PTF and now ZTF, it has become clear that emission as luminous as that accompanying SN\,1998bw is rare.
Without a GRB trigger it is challenging to discover explosions similar to SN\,2006aj, which had a low-frequency radio light curve that peaked within the first five days and faded more quickly than the light curve of SN\,1998bw.
In the case of ZTF18aaqjovh, X-ray observations within the first 10 days may have enabled us to detect an X-ray light curve like that accompanying SN\,2006aj,
but we were unable to observe with Swift due to the proximity of ZTF18aaqjovh to the Sun at the time.

At present, Ic-BL SNe are discovered and classified via brute-force spectroscopy, so unless they are very nearby they are typically not recognized until a week after explosion.
It would be useful to develop strategies for discovering Ic-BL SNe earlier in their evolution, perhaps based on the properties of their host environment,
or --- in higher-cadence surveys -- from the presence of an early ($<1\,\days$) peak in the optical light curve, like that seen in SN\,2006aj and SN\,1998bw.
These could perhaps be distinguished from double-peaked light curves of other SN progenitors (e.g. \citealt{Fremling2019c}) by the luminosity of this first peak, if the redshift to the SN is known.

When the paper has been accepted for publication the data will be made publicly available via WISeREP, an interactive repository of supernova data \citep{Yaron2012}.
The code to produce the figures in this paper has been released under  \dataset[10.5281/zenodo.3634931]{https://doi.org/10.5281/zenodo.3634931}.

\acknowledgements

We would like to thank the anonymous referee for taking the time to do a thorough reading of the manuscript,
and for providing detailed and useful comments.
A.Y.Q.H. thanks Kevin Hurley for maintaining the IPN master
burst list (ssl.berkeley.edu/ipn3/masterli.html) that was used to
compile Table 6,
and Daniel Goldstein and Yuhan Yao (Caltech) for useful discussions regarding ZTF photometry.

Based on observations obtained with the Samuel Oschin Telescope 48 inch and the 60 inch Telescope at the Palomar Observatory as part of the Zwicky Transient Facility project. ZTF is supported by the National Science Foundation under grant No. AST-1440341 and a collaboration including Caltech, IPAC, the Weizmann Institute for Science, the Oskar Klein Center at Stockholm University, the University of Maryland, the University of Washington, Deutsches Elektronen-Synchrotron and Humboldt University, Los Alamos National Laboratories, the TANGO Consortium of Taiwan, the University of Wisconsin at Milwaukee, and Lawrence Berkeley National Laboratories. Operations are conducted by COO, IPAC, and UW.
This work made use of data supplied by the UK Swift Science Data Centre at the University of Leicester.
This work made use of the data products generated by the NYU SN group, and 
released under doi:10.5281/zenodo.58767, 
available at \url{https://github.com/nyusngroup/SESNspectraLib}.
SED Machine is based upon work supported by the National Science Foundation under grant No. 1106171.
The ZTF forced-photometry service was funded under the Heising-Simons Foundation grant \#12540303 (PI: Graham).
Partially based on observations made with the Nordic Optical Telescope, operated by the Nordic Optical Telescope Scientific Association at the Observatorio del Roque de los Muchachos, La Palma, Spain, of the Instituto de Astrofisica de Canarias.
The data presented here were obtained in part with ALFOSC, which is provided by the Instituto de Astrofisica de Andalucia (IAA) under a joint agreement with the University of Copenhagen and NOTSA.
The National Radio Astronomy Observatory is a facility of the National Science Foundation operated under cooperative agreement by Associated Universities, Inc.

A.Y.Q.H. is supported by a National Science Foundation Graduate Research Fellowship under grant No.\,DGE‐1144469. A.C. acknowledges support from the National Science Foundation CAREER award \#1455090. A.C. and A.Y.Q.H. acknowledge support from the Chandra GI award \#19500451.
This work was partially supported by the GROWTH project funded by the National Science Foundation under PIRE grant No.\,1545949.
A.A.M. is funded by the Large Synoptic Survey Telescope Corporation, the Brinson Foundation, and the Moore Foundation in support of the LSSTC Data Science Fellowship Program; he also receives support as a CIERA Fellow by the CIERA Postdoctoral Fellowship Program (Center for Interdisciplinary Exploration and Research in Astrophysics, Northwestern University).
C.F. gratefully acknowledges support of his research by the Heising-Simons Foundation
(\#2018-0907)

\end{document}